\documentstyle[twocolumn,aps]{revtex}

\newcommand{\qv}{\mbox{{\bf q}}}
\newcommand{\kv}{\mbox{{\bf k}}}

\input{epsf}
\begin{document}
\draft
\title{On the relation between the boson-fermion model and RVB state}
\author{Evgueny Kochetov}
\address{Bogolyubov Theoretical laboratory, Joint Institute for
Nuclear Research, 141980 Dubna, Russia}
\author{Marcin Mierzejewski}
\address{Institute of Physics, University of Silesia, 40-007 Katowice, Poland}
\maketitle
\begin{abstract}
We show that
the resonating valence bond (RVB) state of the $t$--$J$ Hamiltonian
can be described within an
effective boson--fermion model.
We derive a boson--fermion Hamiltonian
which describes the lowest order coupling
between holons and spinon waves.
The RVB order parameter determines a hybridization between
the both types of excitations.
The effective Hamiltonian is investigated within the lowest order
self-consistent conserving diagrammatic approximation. We compare our
approach with the phenomenological boson--fermion model (BFM)
that is believed to describe the pseudogap phase.

\end{abstract}

The origin of the  pseudogap is one of
the most important current problems  in
high temperature superconductors.
This normal state gap is seen by
various  techniques, including angle-resolved photoemission
spectroscopy \cite{arpes1,arpes2,arpes3},
NMR\cite{nmr1,nmr2}, infrared \cite{infra}, and
transport\cite{transport} measurements.
Although we now possess   considerable  experimental data,
the understanding of this phenomenon is still far from
complete. There are a number of  microscopic scenarios,
but none of them is  commonly  accepted.
The opening of the pseudogap, and its evolution into
the true superconducting gap, may, however,  be described by
a phenomenological boson--fermion model \cite{robasz,ran95,ran00,dom,romano}.
This BFM is usually understood as a system of localized tightly bound
electron pairs (bosons) which hybridize with itinerant electrons.
This model customarily neglects  the hard core effects of the bound
electron pairs \cite{ran95,ran00,dom} and assumes the standard
commutation relations for the boson operators.
We may then identify three characteristic temperatures \cite{ran00}
which correspond to: ({\it i}) the opening of the pseudogap in the
fermion excitation spectrum, ({\it ii}) the appearance of itinerant bosons,
({\it iii}) the  condensation of these bosons and the
concurrent onset of superconductivity.
Due to the phenomenological character of the BFM, the microscopic
origin of the pair correlations is not specified. Moreover,
it is not straightforward to select an appropriate set of
model parameters ---  in particular the total number of fermions and
bosons. In order to answer these questions we need to  find
a relationship between the BFM and a microscopic approach\cite{auerbach}.
The aim of this work is to provide such a link by
demonstrating a close similarity
between the BFM and the $t$--$J$ model with a mean-field RVB state.

As it is known, the RVB state~\cite{anderson}
is characterized by three important features: ({\it i})
spin--charge separation, ({\it ii})
short range antiferromagnetic (AF) correlations,
and ({\it iii}) holon
condensation. Owing to ({\it i}), the RVB pairing amplitude
may develop at a much higher
temperature than the superconducting transition temperature,
at which the phase coherence between the
electron pairs is established~\cite{kivelson}.
In  the RVB
scenario~\cite{anderson}
there is a phase with the preformed, uncoherent, electron pairs
with an amplitude $\Delta_{{\rm RVB}}$ that serves as the RVB order parameter.
In this phase the RVB electron pairs are to be thought of as spinon singlets
which are coupled to holon excitations, this resulting in a
boson-fermion (BF)
interaction. It is therefore natural to suggest that it is
the RVB order parameter that controls the pseudogap.
This is our basic assumption.

First we demonstrate
that the  RVB state can be described by a
holon--spinon interaction.
Next, application of the
linear spinon--wave approximation leads to a boson--fermion
model where bosons are doubly charged with respect to spinless fermions.

To begin, we express the $t$--$J$ model in terms of the Hubbard operators
$X^{\sigma 0}_i=c^{\dag}_{i\sigma}(1-n_{i,-\sigma})$,
$$H_{t-J}=-t\sum_{\left<i,j,\sigma\right>}X_i^{\sigma 0}X_j^{0\sigma}+
{\rm H.c.}
+J\sum_{\left<i,j \right>}\left(\vec Q_i\vec Q_j-\frac{n_in_j}{4}\right),$$
where $c_{i\sigma}$ annihilates an electron at site $i$
with spin projection $\sigma=\pm, \,n_{i\sigma}\equiv c^{\dag}_{i\sigma}
c_{i\sigma}$, $n_i=\sum_{\sigma} n_{i \sigma}$
and $\sum_{\left<i,j \right>}$ denotes a summation over the nearest neighbor
{\it nonrepeated} bonds.
The electron--spin operator,
$\vec Q_i=\frac{1}{2}\sum_{\sigma\sigma'}X^{\sigma0}_i
\vec{\tau}_{\sigma\sigma'}X^{0\sigma'}_i$,
enters the magnetic part of $H_{t-J}$ and
$\vec{\tau}=(\tau^1,\tau^2,\tau^3)$ stand for the Pauli matrices.
The operators $\vec Q$ fulfil the ${\rm SU(2)}$
algebra commutation relations, but do not form
a true representation, since the
second Casimir operator ${\vec Q}^2$ is not a c-number.

Note that the $X^{\sigma 0}$ appear as fermionic-like
operators, while the
$X^{\sigma\sigma'}=X^{\sigma 0}X^{0\sigma'}$ correspond to bosonic degrees
of freedom.
The crucial  $t$--$J$ model requirement of no double occupancy
$\sum_{\sigma}
c^{\dag}_{i\sigma}c_{i\sigma}\le 1$  follows automatically: by
definition
$X^{00}+\sum_{\sigma}X^{\sigma\sigma}=I.$ Unlike
the slave--particle
representations,  this restriction imposes an additional constraint on
dynamical degrees of freedom and  seems to pose a severe problem
in a mean--field treatment.
Note further that the $t$--$J$ Hamiltonian possesses
two global ${\rm U(1)}$ symmetries:
${\rm U}_{N_e}(1)$ and ${\rm U}_{Q_3}(1)$.
These correspond to the conservation of the total electron number
$N_e=\sum_i(X^{++}_i+X^{--}_i)
=\sum_i n_i$, and to the total spin projection
$Q_3=\frac{1}{2}\sum_i(X^{++}_i-X^{--}_i)$, respectively.

A conventional Hartree-Fock decoupling is applied to the magnetic part of the
$t$--$J$ Hamiltonian
$$H_J:=
J\sum_{\left<i,j \right>}\left(\vec Q_i\vec Q_j-\frac{1}{4}n_in_j\right)
=-J\sum_{\left<i,j \right>}b^{\dag}_{ij}b_{ij}$$
to yield
$$H_J \simeq -J\sum_{\left<i,j \right>}(\Delta_{ij}b^{\dag}_{ij}
+\Delta^*_{ij}b_{ij} -|\Delta_{ij}|^2),$$
where
$b^{\dag}_{ij}=\frac{1}{\sqrt{2}}\left(X^{+0}_iX^{-0}_j-
X^{-0}_iX^{+0}_j\right)$,
is the valence bond "singlet" pair creation operator
and $\Delta_{ij}$ is the
RVB order parameter defined on each bond between the nearest neighbor sites.
$\Delta_{ij}\neq 0$ breaks the global ${\rm U}_{N_e}(1)$ symmetry,
though it does not directly result in superconductivity.
It instead indicates the onset
of the electron spin-singlet formation. There are many possible
mean field states
in RVB theory, among which the extended $s$--phase
is selected here because of its simplicity.

To proceed, we employ the Feynman path--integral representation for
the partition function.
We use  the {\it geometric quantization approach} based on
the Poisson realization of the ${\rm SU(2|1)}$ superalgebra on a
classical phase space ---  the supersphere $CP^{1|1}$~\cite{kochetov}.
In this approach there is an
isomorphic map $X\to X^{cl}$ of the ${\rm SU(2|1)}$
generators into classical functions on the supersphere which
preserves the Lie superalgebra
structure. This map can be realized explicitly by exhibiting covariant symbols
(known as the momentum maps) of the Hubbard
operators in terms of the ${\rm SU(2|1)}$ coherent
states, $|z,\xi\rangle$~\cite{kochetov}:
$X^{cl}=\langle z,\xi|X | z,\xi \rangle$. Then,
$Q^{cl}_3=-\frac{1}{2}(1-|z|^2)w$,
$(Q^{+})^{cl}=zw$,
$(Q^{-})^{cl}=\bar zw$,
$(X^{0-})^{cl}=-z\bar\xi w$,
$(X^{0+})^{cl}=-\bar\xi w$,
$(X^{+0})^{cl}=-\xi w$,
$(X^{-0})^{cl}=-\bar z\xi w$,
$(X^{++})^{cl}+(X^{--})^{cl}=(1+\bar zz)w$,
$(X^{00})^{cl}={\bar\xi}\xi w$, where
$w:=(1+|z|^2+\bar\xi\xi)^{-1}$.
Here the  even $z$ and odd $\xi$ Grassmann variables parameterize superspace
${\rm SU(2|1)}/{\rm U(1|1)}=CP^{(1|1)}$, which is
 the $N=1$ supersymmetric extension of the
Riemannian sphere $CP^1$. These functions form a
${\rm SU(2|1)}$
superalgebra under the Poisson superbrackets on $CP^{1|1}$.
For an arbitrary fermionic Hamiltonian with the
requirement of no double occupancy, the relevant Berezin
path-integral representation of the partition function is equivalent
to the ${\rm SU(2|1)}$ phase-space path integral
$$Z=\int_{CP^{1|1}}
D\mu_{\rm SU(2|1)}\exp\left(\int\langle z,\xi|\frac{d}{dt}-H
|z,\xi\rangle dt\right),$$
$$ D\mu_{\rm SU(2|1)}=
\frac{d\bar zdzd{\bar\xi}d\xi}{2\pi i(1+|z|^2+{\bar\xi}\xi)},
$$
where the Hamiltonian is expressed in terms of the $X$--operators.
Despite a rather complicated appearence,
the ${\rm SU(2|1)}$ path-integral representation for the $t$--$J$ model is
convenient
since  ({\it i}) it automatically incorporates the requirement of no
double occupancy,
$
(X^{00})^{cl}+ \sum_{\sigma}(X^{\sigma\sigma})^{cl}\equiv I
$, independetly of any approximations being applied to $z$ and $\xi$
fields;
 ({\it ii}) it furnishes a convenient set of independent dynamical
variables $(z,\xi)\in CP^{1|1}$,
which can directly be related to the charge and
spin degrees of freedom.

The coordinates $z$ and $\xi$ can be related to the ${\rm SU(2)}$
spinon and ${\rm U(1)}$ holon
degrees of freedom in the following manner: We make two
successive changes of variables
$z\to z\sqrt{1+{\bar\xi}\xi},\,\,\xi\to\xi\sqrt{1+|z|^2}$ to decouple the
${\rm SU(2|1)}$ measure, $D\mu_{\rm SU(2|1)}
\to D\mu_{\rm SU(2)}D\mu_{\rm U(1)}$,
where
$D\mu_{\rm SU(2)}=d\bar zdz/\pi i(1+|z|^2)^2$
stands for the ${\rm SU(2)}$ invariant measure on $CP^1$,
while
$D\mu_{\rm U(1)}=d\bar\xi d\xi$
denotes the Berezin integration over odd Grassmann variables\cite{kochetov1}.

The $t$--$J$ partition function becomes
\begin{eqnarray}
Z_{t-J}&=&\int_{CP^1}\prod_j
D\mu^{(j)}_{\rm SU(2)}\int\prod_jD\mu^{(j)}_{\rm U(1)}\exp[{\cal A}_{t-J}],\nonumber\\
{\cal A}_{t-J}&=&\frac{1}{2}\sum_j\int_0^{\beta}
\left(\frac{\dot{\bar z_j}z_j-\bar z_j\dot z_j}{1+|z_j|^2}+
\dot{\bar\xi_j}\xi_j-\bar\xi_j\dot\xi_j\right)dt\nonumber\\
&-&\int_0^{\beta}{H}^{cl}_{t-J}dt,
\label{eq:2.4}\end{eqnarray}
with $\xi_i$ corresponding to the ${\rm U(1)}$
charged spinless fermion degrees of
freedom (holons),
and $z_i$ representing pure ${\rm SU(2)}$ spins (spinons).

The classical Hamiltonian now reads
\begin{eqnarray}
H^{cl}_{t-J}&=&-t\sum_{\left<i,j \right>}\xi_i\bar\xi_j \langle z_i|z_j\rangle
\nonumber \\
&-&\frac{J\Delta_{\rm RVB}}{2}\sum_{\left<i,j \right>}
\xi_i\xi_j\Phi(\bar z_j,\bar z_i) +{\rm H.c.}\nonumber\\
&-&\mu^{\prime}\sum_i(1-\bar\xi_i\xi_i)-\lambda\sum_i\left[2S_3(\bar z_i,z_i)+
\bar\xi_i\xi_i\right].
\label{eq:2.5}
\end{eqnarray}
We have dropped the constant term. We have also explicitly introduced
a chemical potential term,
$-\mu^{\prime} N_e$, as well as the Langrange multiplier $\lambda$
which controls the magnitude of the total electron magnetic moment.

The ${\rm SU(2)}$ algebra is realized on $CP^1$ in terms of the momentum maps
$S^{cl}_3=-\frac{1}{2}\,\frac{1-|z|^2}{1+|z|^2},\quad
(S^{+})^{cl}=\frac{\bar z}{1+|z|^2},\quad (S^{-})^{cl}
=\frac{z}{1+|z|^2},\quad z\in CP^1$
and
$$\Phi(\bar z_j,\bar z_i)
=\frac{(\bar z_j-\bar z_i)}{\sqrt{(1+|z_i|^2)(1+|z_j|^2)}}=:\Phi_{ij}$$
appears as a spinon--singlet amplitude,
$\Phi_{ij}=$ $\Psi_{s_3=-1/2}(z_i)$ $ \Psi_{s_3=1/2}(z_j)
-$ $\Psi_{s_3=-1/2}(z_j)$ $\Psi_{s_3=1/2}(z_i),$
where $\Psi_{s_3=\pm 1/2}(z)$ is a spinon wave function in the basis of the
${\rm SU(2)}$ coherent states,
$|z\rangle=
(1+|z|^2)^{-1/2}\exp\{z S_{-}\}
|s_3=1/2\rangle.$ Namely,
$\Psi_{s_3=\pm 1/2}(z)=\langle s_3= \pm 1/2|z\rangle,\quad
S_{(3)}|s_3\rangle=s_3|s_3\rangle$
and the $SU(2)$ coherent-state inner product is
$$\langle z_i|z_j\rangle
=\frac{(1+\bar z_iz_j)}
{\sqrt{(1+|z_i|^2)(1+|z_j|^2)}},$$
so that $|\Phi_{ij}|^2=1-|\langle z_i|z_j\rangle|^2$.

In general, the Hamiltonian (\ref{eq:2.5}) cannot be thought of as a covariant
symbol of a Fermi--Bose type operator, except for a particular case of
the linear {\it spinon}--wave (LSW) approximation. This approximation
amounts to expanding the  action~(\ref{eq:2.4}), as well as a
 measure factor in the
path integral, in powers of $|z|^2$ up to the leading order.
In the paramagnetic
phase $\langle Q^{LSW}_{3,i}\rangle=0$, and the spinon fluctuations
become bounded
by the condition $\langle |z_i|^2\rangle \le 1/2$. This,
to some extent, justifies the LSW approximation. Within
this approximation
the partition function, takes on the form
\begin{eqnarray}
Z_{\rm BF}&=&\int \prod_j
d \bar{z}_j d z_j\:
d \bar{\xi}_j d {\xi}_j\:
\exp[{\cal A}_{\rm BF}],\nonumber\\
{\cal A}_{\rm BF}&=&\frac{1}{2}\sum_j\int_0^{\beta}
\left(\dot{\bar z_j}z_j-\bar z_j\dot z_j+
\dot{\bar\xi_j}\xi_j-\bar\xi_j\dot\xi_j
-{H}^{cl}_{\rm BF}\right) dt, \nonumber
\end{eqnarray}
where the classical Hamiltonian reads:
\begin{eqnarray}
H^{cl}_{\rm BF}&=&-t\sum_{\left<i,j \right>}\xi_i\bar\xi_j
-\frac{J\Delta_{\rm RVB}}{2}\sum_{\left<i,j \right>}
\xi_i\xi_j (\bar z_j-\bar z_i) +{\rm H.c.}\nonumber\\
&-&\mu^{\prime}\sum_i(1-\bar\xi_i\xi_i)-\lambda\sum_i
\left[-1+2 \bar z_i z_i+\bar\xi_i\xi_i\right]. \nonumber
\end{eqnarray}
It straightforward to see that the above partion function
represents the following  boson--fermion system:
\begin{eqnarray}
H_{BF}&=&-t\sum_{\left<i,j \right>}f_if^{\dagger}_j-2v
\sum_{\left<i,j \right>}f_if_j(b^{\dagger}_j-b^{\dagger}_i)+H.c.\nonumber\\
&+& \Delta_B\sum_ib^{\dagger}_ib_i -\mu\sum_i(2b^{\dagger}_ib_i+
f^{\dagger}_if_i),\nonumber\\
&&\{f_i,f^{\dagger}_j\}=[b_j,b^{\dagger}_j]=\delta_{ij}.
\label{eq:2.6}\end{eqnarray}
The parameters of the boson--fermion system
are expressed in terms of the parameters of the original $t$--$J$ Hamiltonian
and the RVB state:
$ v=J \Delta_{\rm RVB}/4$; $ \Delta_{B}=2 \mu^{\prime}$;
$\mu=\mu^{\prime}+\lambda$.
The ${\rm U}_{Q_3}(1)$ symmetry of eq.~(\ref{eq:2.5}) under global gauge
transformations $z_i\to e^{2i\theta}z_i,\, \xi_i \to e^{i\theta}\xi_i$
leads to conservation of the total charge
$\sum_i(2b^{\dagger}_ib_i+f^{\dagger}_if_i)$.
Note, that this global ${\rm U(1)}$ symmetry characterizes the standard BFM
and indicates that bosons are doubly charged with
respect to fermions. The BFM discussed in Refs. \cite{ran95,ran00,dom}
is refered to as the standard approach.
One should also keep in mind
that this symmetry is an intimate feature of
the RVB state, and
occurs independently of the LSW aproximation.

The standard BFM was investigated with the help of various
methods, e.g.: the lowest order self--consistent conserving
diagrammatic approximation \cite{ran95,ran00}, the dynamical mean--field
approximation \cite{romano} and the continuous unitary transformation
\cite{dom}. Unlike the standard BFM, the Hamiltonian obtained
in the present study involves spinless fermions.
Therefore decay and recombination between fermions and bosons
involves two neighboring lattice sites.
In the present case, bosons with momentum
$\qv=0$ do not participate in these processes,
in the
standard BFM these bosons are of crucial importance.
Because of   these differences, we cannot directly adopt the
numerical results obtained for the standard BFM.
Therefore, we have investigated a one dimensional (1D) case within
the lowest order self--consistent conserving diagrammatic approximation
and compared our results to those presented in Refs.\cite{ran95,ran00}.

In the the lowest order self--consistent conserving diagrammatic approximation
one obtains a system of equations for the fermion,
$G_F(\kv, \omega)$, and boson, $G_B(\kv, \omega)$,
Green's function. We refer to Ref. \cite{ran00} for the details.
This set of equations has been solved iteratively
slightly above the real axis, i.e. for $\omega= {\rm Re} \omega +i0.01t$.
We have used grids of 501 momenta in the interval $[0,\pi]$,
and $2001$ frequencies in the interval $[-5t,5t]$. We have also
fixed the bosonic level $\Delta_B=-2.4t$.

\begin{figure}
\epsfxsize=7cm
\centerline{\epsffile{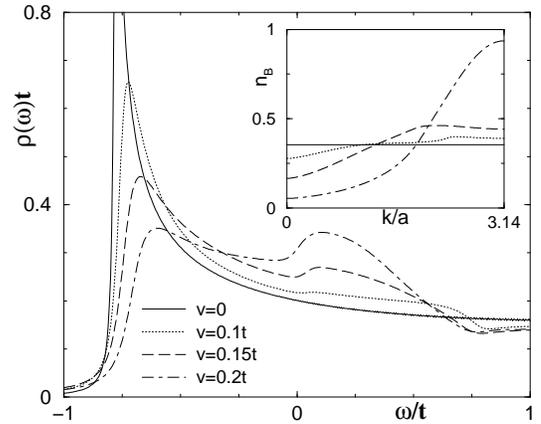}}
\caption{The fermionic density of states calculated at the temperature
$kT=0.02t$ for  different values of the boson--fermion coupling.
$\omega=0$ corresponds the Fermi level.
The inset shows the average bosonic occupation number
as a function of momentum.}
\end{figure}

Fig. 1 shows the fermionic density of states, $\rho(\omega)$,
calculated for different values of the  RVB order parameter.
The boson--fermion coupling enhances
the density of states just above the Fermi level and reduces
$\rho$ for much larger energies.
Effectively, the local minimum in the
density of states occurs at the Fermi level.
Modification of the density of states takes place also for a weak
boson--fermion coupling, when the local minimum at the Fermi level
is almost not visible. Therefore, despite the absence of the gap
the system cannot be considered as a composition of noninteracting
fermions and bosons. The most important difference between the
standard BFM and the present study shows up in the average
boson occupation number $n_B$ (see the insets in Figs. 1 and 2).
Here, the
boson--fermion coupling enhances the number of bosons with
momentum close to $\pi$, whereas in the standard BFM the bosons
with momentum $\qv=0$ play the dominating role.
In the present case
the momentum dependence of $n_B$ can be attributed
to the AF short range correlations.
However, bosons are uniformly distributed in the real space
and there is no antiferromagnetic order (in agreement with the assumption
of the RVB state).

\begin{figure}
\epsfxsize=7cm
\centerline{\epsffile{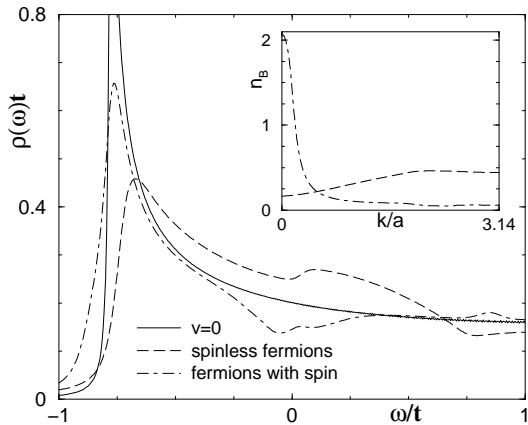}}
\caption{The fermionic density of states calculated
for ({\it i}) non interacting particles,
({\it ii}) standard BFM (fermions with spin)
and ({\it iii}) obtained in the present study (spinless fermions).
The inset shows the average boson occupation number
as a function of momentum in the cases ({\it ii}) and ({\it iii}).}
\end{figure}

In Fig. 2 we compare $\rho(\omega)$ and $n_B$
obtained within the standard BFM and the present study.
In the case of spinless fermions $n_B(\kv)$ is relatively flat,
when compared to the standard approach. It
suggests that bosons do not condensate at the state with
some particular momentum. However, investigation of this problem
is not possible within a one--dimensional numerical study.
One should also keep in mind, that
in our approach (which concerns only the mean field RVB state)
condensation of bosons does not represent the superconducting
phase transition. Similarly to the standard BFM, the
local minimum in $\rho$, that occurs at the Fermi level,
deepens with the decreasing temperature and disappears when
the temperature is sufficiently high (see Fig. 3).

The difference between the present approach and
the standard BFM shows also up in the fermionic
spectral functions (see Fig. 4). Here, the most
important broadening of the spectral functions
takes place above and below the Fermi surface.
In the standard case \cite{ran95} the
spectral functions have a three--peaked
structures, which are mostly visible
at the Fermi surface. It may be attributed
to the free-particle--like motion
of the bosons with small wave vectors.
However, in the present approach these bosons
are not involved in the decay and recombination
processes.

\begin{figure}
\epsfxsize=7cm
\centerline{\epsffile{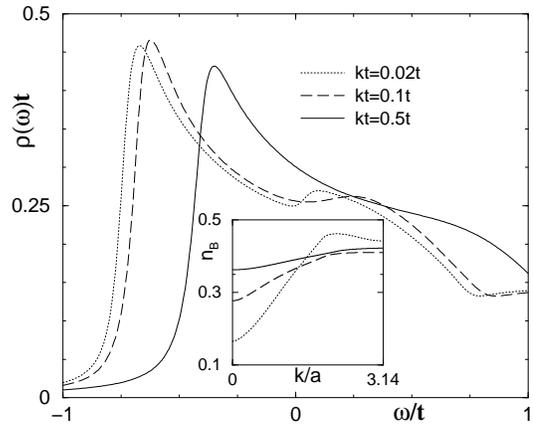}}
\caption{The same as in Fig. 1 but calculated at different temperatures
for $v=0.15t$.}
\end{figure}

To conclude, starting from  an RVB state for
the $t$--$J$ Hamiltonian, we have derived a connection
between the $t$--$J$ model and the BFM.
We have expressed the parameters of the effective
Hamiltonian in terms of those of
the $t$--$J$ model and the RVB state. The resulting
model differs from the standard one in that
the fermionic degrees of freedom correspond to
spinless quasiparticles.
However, as in   the standard BFM, a local minimum
in the fermionic density of states
occurs close to the Fermi level.
The minimum deepens with the decreasing of temperature
and vanishes when the temperature is sufficiently high.
The present study is restricted only to the
mean--field RVB state,
and does not attempt to account for the superconducting phase transition.
It
is, however, free from the ambiguities which occur in the
phenomenological BFM when tries to select an appropriate set of model parameters.
\begin{figure}
\epsfxsize=7cm
\centerline{\epsffile{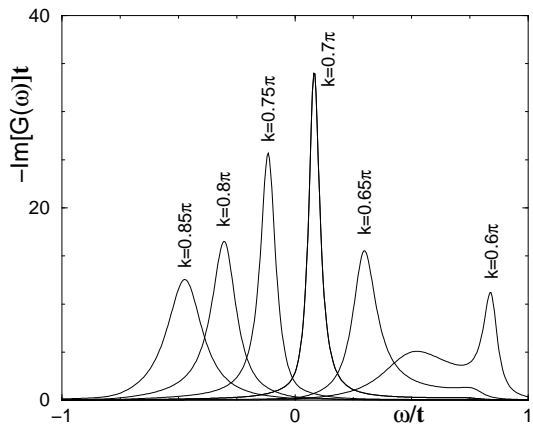}}
\caption{Spectral function for fermions
calculated for wave--vectors, which are in the
vicinity of the Fermi surface. We have used $v=0.15t$ and
$kT=0.02t$.}
\end{figure}

\end{document}